\documentclass[letterpaper, 10 pt, conference]{ieeeconf}  

\IEEEoverridecommandlockouts                              
\usepackage{xurl}
\usepackage[utf8]{inputenc}
\usepackage[T1]{fontenc}
\usepackage{graphicx}
\usepackage{booktabs}
\usepackage{threeparttable}
\usepackage{graphics} 
\usepackage{epsfig} 
\usepackage{mathptmx} 
\usepackage{mathptmx} 
\usepackage{amsmath} 
\usepackage{amssymb}  

\title{\LARGE \bf
A Two-sided Model for EV Market Dynamics and Policy Implications
}

\author{Haoxuan Ma$^{1}$, Brian Yueshuai He$^{1*}$, Tomas Kaljevic$^{2}$ and Jiaqi Ma$^{1}$
\thanks{$^{1}$Department of Civil and Environmental Engineering, Samueli School of Engineering, University of California at Los Angeles, Los Angeles, CA 90095}
\thanks{$^{2}$Department of Computer Science, Samueli School of Engineering, University of California at Los Angeles, Los Angeles, CA 90095}
\thanks{$^{*}$Corresponding Author: Brian Yueshuai He (e-mail: yueshuaihe@ucla.edu)}%
}

\begin{document}

\maketitle
\thispagestyle{empty}
\pagestyle{empty}

\begin{abstract}

The diffusion of Electric Vehicles (EVs) plays a pivotal role in mitigating greenhouse gas emissions, particularly in the U.S., where ambitious zero-emission and carbon neutrality objectives have been set. In pursuit of these goals, many states have implemented a range of incentive policies aimed at stimulating EV adoption and charging infrastructure development, especially public EV charging stations (EVCS). This study examines the indirect network effect observed between EV adoption and EVCS deployment within urban landscapes. We developed a two-sided log-log regression model with historical data on EV purchases and EVCS development to quantify this effect. To test the robustness, we then conducted a case study of the EV market in Los Angeles (LA) County, which suggests that a 1\% increase in EVCS correlates with a 0.35\% increase in EV sales. Additionally, we forecasted the future EV market dynamics in LA County, revealing a notable disparity between current policies and the targeted 80\% EV market share for private cars by 2045. To bridge this gap, we proposed a combined policy recommendation that enhances EV incentives by 60\% and EVCS rebates by 66\%, facilitating the achievement of future EV market objectives.

\end{abstract}

\section{INTRODUCTION}

Many countries aim for carbon neutrality and zero emissions by mid-century \cite{c1}. The transportation sector contributes 23\% of global and 29\% of US energy-related emissions \cite{c1}. To meet carbon neutrality, policies to reduce transportation emissions are in place. The US targets zero emissions by 2050 by transitioning to zero-emission vehicles (ZEV) \cite{c3}. Transitioning from traditional internal combustion engine (ICE) vehicles to electric vehicles (EV) is crucial for reducing greenhouse gas (GHG) emissions. Over the past decade, the global EV stock increased from 18 thousand in 2012 to 26 million in 2022 \cite{c1}. This growth is driven by factors such as lower carbon footprint, advancements in mobility technology like connected autonomous vehicles (CAV), intelligent transportation systems (V2X), smart intersections, and economic benefits on fuel costs, especially with rising oil prices \cite{c5, heOJITS2024, c6, shiOJITS2024, devaillyOJITS2024, c7, c8, kesslerOJITS2024}. Public transportation has also progressed in electrification with advancements in routing, scheduling, and charging behavior \cite{HoppeOJITS2023, GalloOJITS2023}. From 2013 to 2019, EV technology advancement led the average EV range to increase from 136 mi to 241 mi \cite{c15}. Meanwhile, battery costs dropped 80\%, battery capacity doubled, and powertrain performance improved \cite{c16}.

To illustrate government policy adjustments for the growing EV market, consider California. Policies like tax rebates and discounts have been implemented to boost EV purchases \cite{c11}. In 2019, California's Clean Cars 4 All program offered up to \$12,000 off for replacing old cars with ZEVs \cite{c11}. In 2021, the California Air Resources Board mandated that by 2035, all new cars sold must be zero-emission vehicles \cite{c12}.

Technological advancements and policy incentives have made EVs more affordable, but factors like charging behavior and infrastructure are also crucial for adoption \cite{c18}. Home chargers are main for private EVs, but public EVCS help reduce range anxiety. Expanding public charging networks addresses current EV owners' needs and promotes wider adoption by easing range concerns \cite{c20}. Studies in the US (2011-2013) and Norway (2010-2015) found charging availability more important than ownership cost \cite{c21, c22}.

Recognizing the impact of EVCS availability on EV adoption, the global market has responded quickly. By the end of 2022, there were 2.7 million public EVCS worldwide, with over 90,000 built that year \cite{c24}. The US government invested \$2.5 billion to increase EV rebates and fund public charging infrastructure \cite{c25}. California's investment in fast chargers and rebate programs like CALeVIP has been crucial in reducing range anxiety and supporting the growing number of EVs \cite{c26}. In summary, as EV adoption grows, so does the demand for public EVCS. Increased public EVCS availability in urban areas helps solve range anxiety, potentially boosting EV adoption. Thus, there is a reciprocal relationship between EV adoption and public EVCS availability, known as the indirect network effect. The contributions of this paper are threefold:
\begin{itemize}
\item \textbf{Modeling indirect network effect of the two-sided EV market}: We present a model that considers the effects of a two-sided market, capturing the indirect network effect between EV adoption and EVCS deployment. This model suggests a convergence state—a point in time when the EV market becomes saturated under current policy levels. 
\item \textbf{Evaluation of a specific EV market and incorporating socio-economic heterogeneity}: We gather real data and apply our model to a metropolitan market with high socio-economic diversity, LA County, which is known for its pioneer position of EV adoption. Our model is at the zip code level to reflect how different socio-demographic factors influence the EV market. We also conduct sensitivity analysis and scenario evaluations to reveal insights into the EV market dynamics of urban settings. 
\item \textbf{Policy recommendation for decision-makers}: We explore how various supply and demand policies can disrupt this equilibrium and lead to a new state of balance. This study can guide policymakers in creating reasonable incentive policies for achieving future transportation electrification goals.
\end{itemize}

\section {Literature Review}

The indirect network and two-sided market effects have been well studied. Katz and Shapiro (1985) first proposed the indirect network effect, defining it as user growth influenced by the adoption of complementary products \cite{c27}. Rochet \& Tirole (2003) and Armstrong (2006) expanded this by distinguishing between one-sided and two-sided markets, focusing on pricing and market strategies that leverage two-sided effects \cite{c28, c29}.

As EVs gain attention, research has studied factors affecting their adoption. Zheng et al. (2022) found financial incentives crucial for boosting EV diffusion \cite{c30}. Technology advancements, especially in battery management, also help overcome adoption barriers like limited range \cite{c31}. Singh (2022) emphasized tailoring incentive programs to socio-demographic patterns for better policy effectiveness \cite{c32}. Increased EV adoption raises demand for public charging stations, which helps alleviate range anxiety, a key barrier identified by Noel et al. (2019) \cite{c33}. Neubauer and Wood (2014) noted that public charging greatly increases utility for drivers, reducing range anxiety \cite{c34}.

Narasimhan et al. (2018) used lagged regression to show how last year's infrastructure impacted this year's EV sales at a state level \cite{c20}. Li et al. (2017) used models of EV demand and EVCS to describe their relationship in US cities between 2011 and 2013 \cite{c21}. Yu et al. (2015) validated Li et al.'s model \cite{c35}. Springel focused on national incentives for EV and EVCS in Norway from 2010 to 2015 \cite{c22}. Xing et al. (2023) analyzed the varied effects of network impacts on vehicle qualities in China \cite{c36}. All papers highlight that the impact of EVCS on EV adoption is as significant as direct factors like price incentives.

Existing literature examines national and regional incentives and technological advancements but lacks studies on localized indirect network effects. The rapid EV market growth from 2019 to 2023, especially in California, highlights the need for detailed studies \cite{c37}. Metropolitan areas like New York City or LA County offer unique EV ecosystems with multi-tiered policy incentives.

Previous studies lack recent, localized data on urban EV market dynamics. Tsukiji et al. (2023) highlight a disparity between EV and EVCS adoption, emphasizing the need to address socio-demographic inequity \cite{c38}. This study examines zip code level market saturation, socio-demographic factors, and policy incentives across government levels to understand community responsiveness and model future EV purchase decisions.

\section{Methodology}

\subsection{Modelling}

Literature often models the indirect two-sided market on EV dynamics nationally and at the state level. Li et al. identify a feedback loop where charging infrastructure boosts EV adoption, which then promotes more infrastructure \cite{c21}. Our model examines the interaction between EVs and EVCS with real historical incentives with a higher granularity, focusing on the socioeconomic differences across zip codes. This reveals how policies and technology can address each neighborhood's needs, promoting a fair transition to electric mobility. We propose a theoretical two-sided market model incorporating critical indirect network effects to understand consumer behavior and EV market evolution. We then apply this framework to historical data, analyzing how elasticities between key interactive and auxiliary terms have contributed to the two-sided market's growth.

Our model contains two key components: the EV demand model and the EVCS supply model. For the EV demand model, we define $z$ as the zip code region and $t$ as the year. The equation is as follows:

\begin{equation}
    \log(s_{zt}) = f\left(\log(E_{zt}), \log(\mathbf{A}_{zt}), \log(\mathbf{T}_{t})\right)
    \label{eq:(1)}
\end{equation}

We define $f()$ as a linear function representing the log-log transformation linear regression equation. $s_{zt}$ is the total EV sales in a certain zip code at a specific time stamp. $E_{zt}$ is the predicted number of cumulative EVCS counts in $z$ at time $t$, which reflects the indirect network effect about how the level of EVCS at a zip code affects that region’s EV adoption. $\mathbf{A}$ is a vector of time and spatial variant auxiliary components such as socio-demographic factors across a $z$.  $\mathbf{T}$ represents other variables that account for the variation, such as oil price and rebate incentives.

\begin{equation}
    \log(E_{zt}) = h\left(\log(Q_{zt}), \log(\mathbf{m}_{zt})\right)
    \label{eq:(2)}
\end{equation}

Correspondingly, as the other part of market modeling, we define $h()$ as a linear function also with a log-log transformation to measure the cumulative sum of EVCS at zip code regions. $Q_{zt}$ represents the cumulative EV install base, the major interaction term between EVCS and EV. 

We can express this term as $Q_{zt} = s_{zt} + \delta Q_{zt-1}$, where $\delta$ is the percentage of EVs operating from $t-1$ to $t$ and $\mathbf{m}_{zt}$ is a vector of time-variant variables that affect the supply side dynamics, such as the saturation level of a certain zip code market. 

To reflect the feedback loop formed by the interaction of both models, assuming we hold other variables as constant $c$, we can expand Equation~\eqref{eq:(1)} and Equation~\eqref{eq:(2)} as:

\begin{equation}
    \log(s_{zt}) = \beta_1\log(E_{zt}) + c_{ev}
    \label{eq:(3)}
\end{equation}
\begin{equation}
    \log(E_{zt}) = \alpha_1\log(Q_{zt}) + c_{evcs}
    \label{eq:(4)}
\end{equation}

Plugging Equation~\eqref{eq:(4)} to Equation~\eqref{eq:(3)}, we can derive the function representing the system dynamics.

\begin{equation}
    \log(s_{zt}) - \beta_1\alpha_1\log(s_{zt}+\delta Q_{zt-1}) = c
    \label{eq:(5)}
\end{equation}

Here, $c$ is a constant term representing all other variables held unchanged. The recursive nature of this function allows us to simulate the indirect network effect in the EV dynamics evolution toward a convergence state. Taking off the log and generalizing it we can express:

\begin{equation}
     s_{t} = \exp(c + \beta_1\alpha_1\log(s_{t}+\delta Q_{t-1}))
     \label{eq:(6)}
\end{equation}

By adjusting the value of $c$, we can study how policy incentive changes can impact the EV market dynamics. Later on, we will run simulations to test different policy scenarios based on this equation.

\subsection{Empirical Framework}

\subsubsection{EV Demand}

Based on our theoretical modeling, we expand it to empirical models that incorporate real historical data. According to Equation~\eqref{eq:(1)}, we expand the following by incorporating real-world factors:

\begin{equation}
   \log(s_{zt}) = \beta_1\log(E_{zt}) + \beta_2\log(B_{zt}) + \beta_3 \log(\mathbf R_{zt}) + \mathbf T_t 
\end{equation}

where $s_{zt}$ is the total EV sales in zip code region $z$ at time $t$. $E_{zt}$ is the predicted cumulative EVCS count in zip code $z$ at time $t$. The term $B_{zt}$ is an important qualifier that measures the burden of purchasing an EV for households in zip code $z$ at time $t$, which we can express as $B_{zt} = \frac{AP_{zt}}{I_{zt}}$. We denote $AP_{zt}$ as the aggregated average price of EVs in zip code $z$ to represent the purchasing power and neighborhood social influence on the acceptable purchase price range. We divide this price by $I_{zt}$, which is the median household income of zip code $z$. Thus, $B_{zt}$ indicates the percentage of annual household income required to purchase an EV in zip code $z$, assuming that a neighborhood shares the same purchase preference and similar income level. According to Figure \ref{fig:ev_sales_forecast}, we suspect the variation of median income level across zip codes can account for different levels of EV stock. Thus, instead of assigning a constant EV price, $B_{zt}$ can better measure EV affordability. $\mathbf R_{zt}$ represents a vector of variables of other region-variant variables including demographic features such as racial/ethnic group. $\mathbf T_t$ is a vector of other time-variant variables such as oil price.

As in the theoretical model mentioned, in Equation~\eqref{eq:(3)}, $E_{zt}$ is the predicted cumulative EVCS count from Equation~\eqref{eq:(2)}, while in Equation~\eqref{eq:(2)}, we understand that the number of EVCS is subject to change based on the cumulative EV install base. $Q$ and $s$ are strongly linear correlated. As a result, this creates a problem of endogeneity. Inspired by Li et al., we introduce the instrumental variable (IV) and a two-step OLS to alleviate the endogeneity problem \cite{c21}. The IV should be correlated with the endogenous variable, EVCS stock, but not related to the dependent variable, which is EV sales. We introduce an IV: the number of commercial parking lots per zip code. This represents potential charging infrastructure development, as more parking lots offer more spaces for charging stations, impacting EV stock. Parking lot numbers don't directly influence EV purchasing decisions. To introduce time variation reflecting EVCS investment trends, we use the lagged total of charging stations outside the current zip code, interacting with parking lot numbers. This accounts for the relative stability of parking lot numbers over yearly periods.

\subsubsection{EVCS}

We expand the EVCS model based on Equation~\eqref{eq:(2)}:

\begin{equation}
    \log(E_{zt}) = \alpha_1 \log(Q_{zt}) + \alpha_2P_{zt} + \alpha_3S_{zt} + \alpha_4I_t
\end{equation}

This model incorporates potential variables that correlate with EVCS growth to fit $E_{zt}$, which is the total number of chargers in zip code $z$ at time $t$.  $Q_{zt}$ is the predicted number of total EVs from stage 1 IV regression. $ P_{zt}$ is the interaction term of the parking lot with the lagged station. $I_t$ represents the percentage of the rebate amount towards the total installation and purchase cost of an EVCS.

However, since private EVCS network providers are dominant in building EV infrastructure, to better capture the supply side incentives, we identified the level of market saturation given zip code as an important factor that impacts the growth of EVCS \cite{c39}. To measure the saturation level given any zip code region, we propose $S_{zt} = \min\max \Bigl( \frac{\log(Q_{zt-1})}{\log(E_{zt-1})} \Bigr)$. We calculate a ratio of EVs per EVCS in a region using the log term of lagged EV stock from the previous year divided by the total EVCS. This ratio is then normalized to a 0-1 range, defining it as a level of saturation. This term captures comparative growth dynamics between EV sales and charging stations on a logarithmic scale. A higher logarithmic value for total EV compared to EVCS indicates faster EV adoption, creating a demand gap for public chargers in a region. This incentivizes private EVCS network providers to build more charging stations.

\section{Experiment and Policy Implication}

\subsection{Data}

Our model utilizes yearly EV sales by type, total EV count, state-level oil prices, EV charger installations, and zip code-level socio-demographic data from open sources like the Department of Energy and AFDC. While applicable to any market with available state data, this study focuses on LA County, California. Yearly new ZEV sales data by fuel type and model per zip code come from the California Energy Commission \cite{c40}. EV stock is derived from Vehicle Fuel Type Count at the California Open Data Portal \cite{c41}. EVCS data, including charging station types and installation years, is from AFDC \cite{c42}. We focus on Level 2 chargers (83\% adoption rate) installed after 2018 to reflect representative charging behavior.

\begin{figure}[htpb]
  \centering
  \includegraphics[width=\columnwidth]{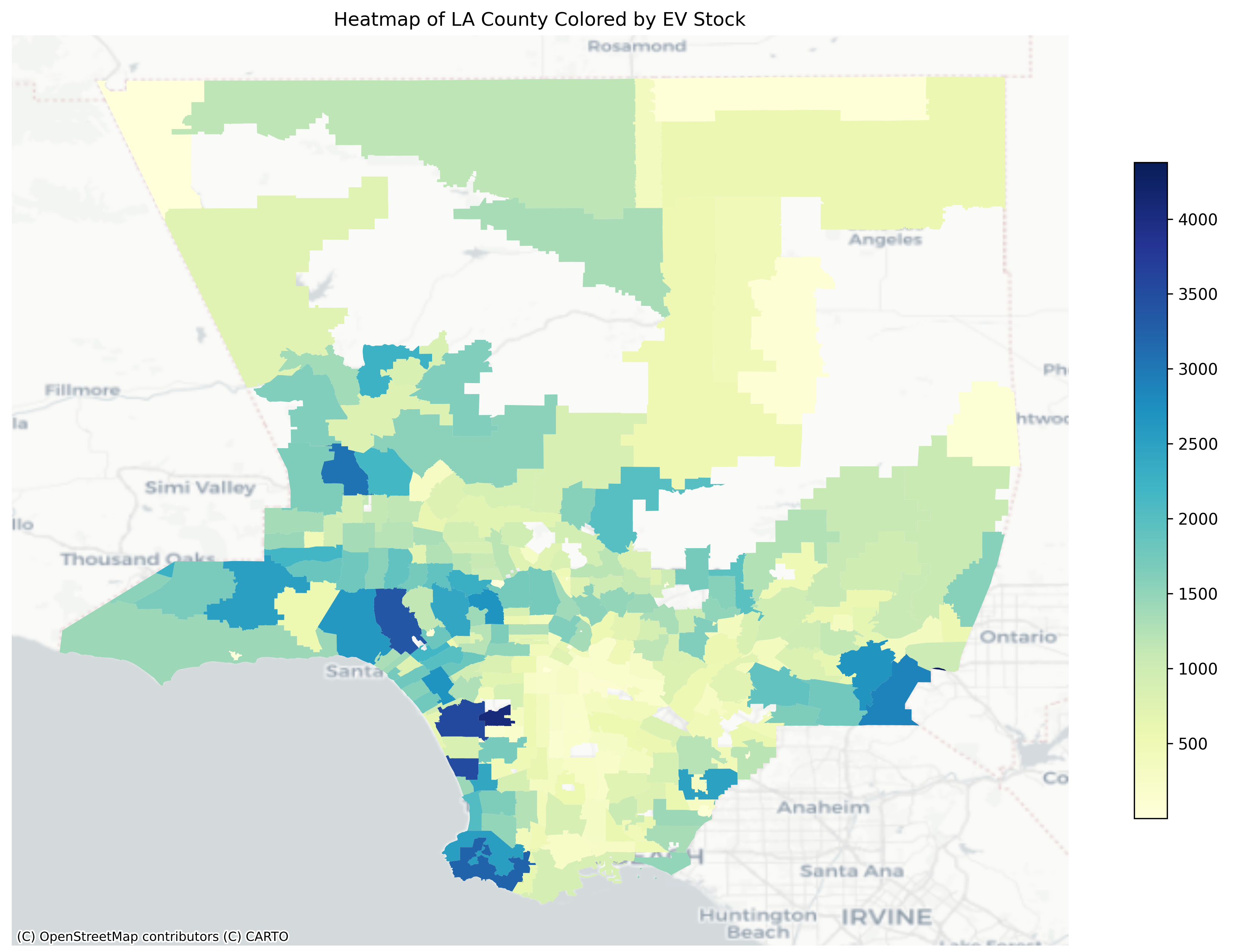}
  \caption{Distribution of EV Stock in LA County in 2022}
  \label{fig:ev_sales_forecast}
\end{figure}

We gathered zip code-level socio-demographic data for LA County from the US Census Bureau, including population, racial distribution, employment rate, housing information, and median household income, to study diverse demographic responses to EV policies and EVCS availability \cite{c43}. Figure \ref{fig:ev_sales_forecast} shows significant regional differences in key statistics. AFDC reports Level 2 EV charger costs decreased from \$7,500 to \$6,000 per unit between 2018 and 2020 \cite{c44, c45}. Assuming linear price reduction, we estimate charger price changes over time.

\subsection{Model Estimation Results}

With the processed data and the empirical model, we employ a 2-stage IV experiment using the Python GMM package to model the dynamics of EV demand, addressing the endogeneity issues arising from the simultaneity of charging station stock and unobserved shocks that may affect the growth of charging stations. To test the effectiveness of GMM, the EV demand model is supplemented by OLS (Ordinary Least Squares) for baseline comparisons. 

\begin{table}[ht]
\centering
\begin{threeparttable}
\caption{Regression results of EV Demand}
\label{tab:regression_results1}
\begin{tabular}{lcc}
\toprule
Variable & GMM & OLS \\
\midrule
ln(Charging station) & 0.3583$^{*}$ (0.0285) & 0.4059$^{*}$ (0.0285) \\
ln(oil\_price)       & 0.8546$^{*}$ (0.0599) & 0.7325$^{*}$ (0.0599) \\
ln(White Population) & 0.1260$^{*}$ (0.0309)  & 0.1188$^{*}$ (0.028) \\
ln(Asian Population) & 0.2592$^{*}$ (0.021)  & 0.2649$^{*}$ (0.016) \\
EV\_Burden           & -3.2022$^{*}$ (0.3368) & -3.4277$^{*}$ (0.178) \\
\midrule
\textit{Summary of Statistics} \\
Number of observations & \multicolumn{1}{c}{} & 996 \\
R$^2$                  & 0.62 & 0.60 \\
\bottomrule
\end{tabular}
\begin{tablenotes}
\small
\item Note: The target variable is $\log(S_{zt})$. Standard error is in the parenthesis.
\item * p-value $\le$ 0.05.
\end{tablenotes}
\end{threeparttable}
\end{table}

Given the log-log formulation of our model, our analysis with GMM can directly assess the elasticity of EV demand in response to percentages of changes in EVCS stock and other factors. The overall R-squared value reached 0.62. This is a reasonable number in the case of IV regression since the IVs prioritize addressing endogeneity over maximizing explained variables. The result from GMM indicates that a 1\% increase in public charging stations leads to a 0.36\% increase in EV demand, indicating a significant positive correlation between EVCS stock and EV sales. Compared with the baseline OLS model, the GMM has a roughly 0.05\% lower correlation between EVCS and EV sales, given the same 1\% increase in EV sales. This indicates that EVCS stock is negatively correlated with unobserved market shocks. We can interpret this as uncaptured market dynamics that hinder EVCS growth; for example, policies like free home charger installation when purchasing EVs that the dealer provides, with higher availability of home chargers, will decrease the market demand for EVCS. 
Oil prices show a robust correlation with ZEV sales trends, with a 1\% increase in oil price leading to a 0.8\% increase in EV sales. This suggests rising oil prices shift consumer preference towards sustainable vehicle options. Socio-demographic features, such as racial group weight by zip codes, also show statistical significance, highlighting the intercorrelation between community characteristics and vehicle adoption patterns.

Another critical aspect, EV burden, varies greatly across different socio-demographic regions. JENI (Justice Equity Need Index) evaluates socio-demographic disparities to address inequities in underdeveloped LA County communities. One of the indicators, the Inequity Driver, measures root inequities across communities that contribute to racial and economic disparities. A lower percentile of Inequity Driver means the region is more underdeveloped. In the case of LA County, the lowest 10 zip codes by Inequity Driver in the year 2022 have an average EV burden of 0.485, whereas for the highest 10 zip codes, the value is 0.158. The great difference suggests how the model can catch the nuances in different socio-demographic regions. Table \ref{tab:regression_results1} demonstrates a strong negative elasticity between EV burden and sales: for every 1\% increase in EV burdens, there is a corresponding 3\% decrease in EV sales. This relationship illustrates the market's sensitivity to EV price volatility and the household financial health in communities. The strong correlation implies to policymakers the impact of social influence and household income conditions within communities on EV adoptions. For instance, adjusting EV rebates based on the poverty level or having special incentives for disadvantaged communities could lead to a more engaged community response, pushing for quicker adoption of EVs.

\begin{table}[ht]
\centering
\begin{threeparttable}
\caption{Regression Result of EVCS}
\label{tab:regression_results2}
\begin{tabular}{lccc}
\toprule
\textbf{Variable} & \textbf{GMM} \\
\midrule
ln(EV Stock) & 0.4992$^{\ast}$ (0.173) \\
ln(parking lot) & 0.0099$^{\ast}$ (0.0599) \\
Saturation & 3.6213$^{\ast}$  (0.104) \\
Rebate Percentage & 1.771$^{\ast}$ (0.54) \\
\midrule
\textit{Summary of Statistics} \\
Number of observations & \multicolumn{1}{c}{996}  \\
R$^2$  & \multicolumn{1}{c}{0.649} \\
\bottomrule
\end{tabular}
\begin{tablenotes}
\small
\item Note: The target variable is $log(E_{zt})$. Standard error is in the parenthesis.
\item * p-value $\le$ 0.05.
\end{tablenotes}
\end{threeparttable}
\end{table}

We employed a GMM model with IV for the EVCS supply analysis. All variables show positive, significant correlations with EVCS numbers. Our interaction term, $Q_{zt}$, shows a strong positive correlation, resulting in a 0.48\% increase in EVCS per 1\% increase in EV stock, reflecting the supply-side response to EV ownership. The saturation level $S_{zt}$ also shows a strong positive coefficient; for each 1\% that EV adoption growth outpaces EV infrastructure growth, charging stations are expected to grow by 3\%. Market saturation in the lowest 10 Inequity Driver zip codes is 42\% higher than in the highest 10, indicating growth potential in underserved areas. Financial incentives ($I_t$) are effective; a 1\% increase in charger cost rebate leads to a 1.77\% increase in EVCS numbers. These findings demonstrate how EV adoption, market saturation, community inequity, and financial incentives drive EVCS deployment.

\subsection{Policy Implications}

Our EV demand and EVCS model captures indirect network effects. As a long-term goal, the National Renewable Energy Lab (NERL) and LA Department of Water and Power (LADWP) aim for 80\% of vehicles in LA County to be ZEVs by 2045 \cite{c47}. By incorporating real policy incentives, our model predicts the future EV market in LA County under current policies, assessing if they are sufficient for the NERL and LADWP goal. We first predict EV sales in LA County over the next 30 years with current incentives and then evaluate different scenarios to determine the needed incentives to reach 80\% EVs by 2045.

Assuming private cars remain the primary travel mode, we consider vehicle numbers in LA County proportional to population. We trained a linear model using historical population data from the California Department of Finance and vehicle sales data from the California DMV (2002-2022) to predict future vehicle counts based on population projections \cite{c48}. The model predicts 6.22 million vehicles in LA County by 2045, with estimated annual sales of 437,000.

\begin{figure}[htpb]
  \centering
  \includegraphics[width=\columnwidth]{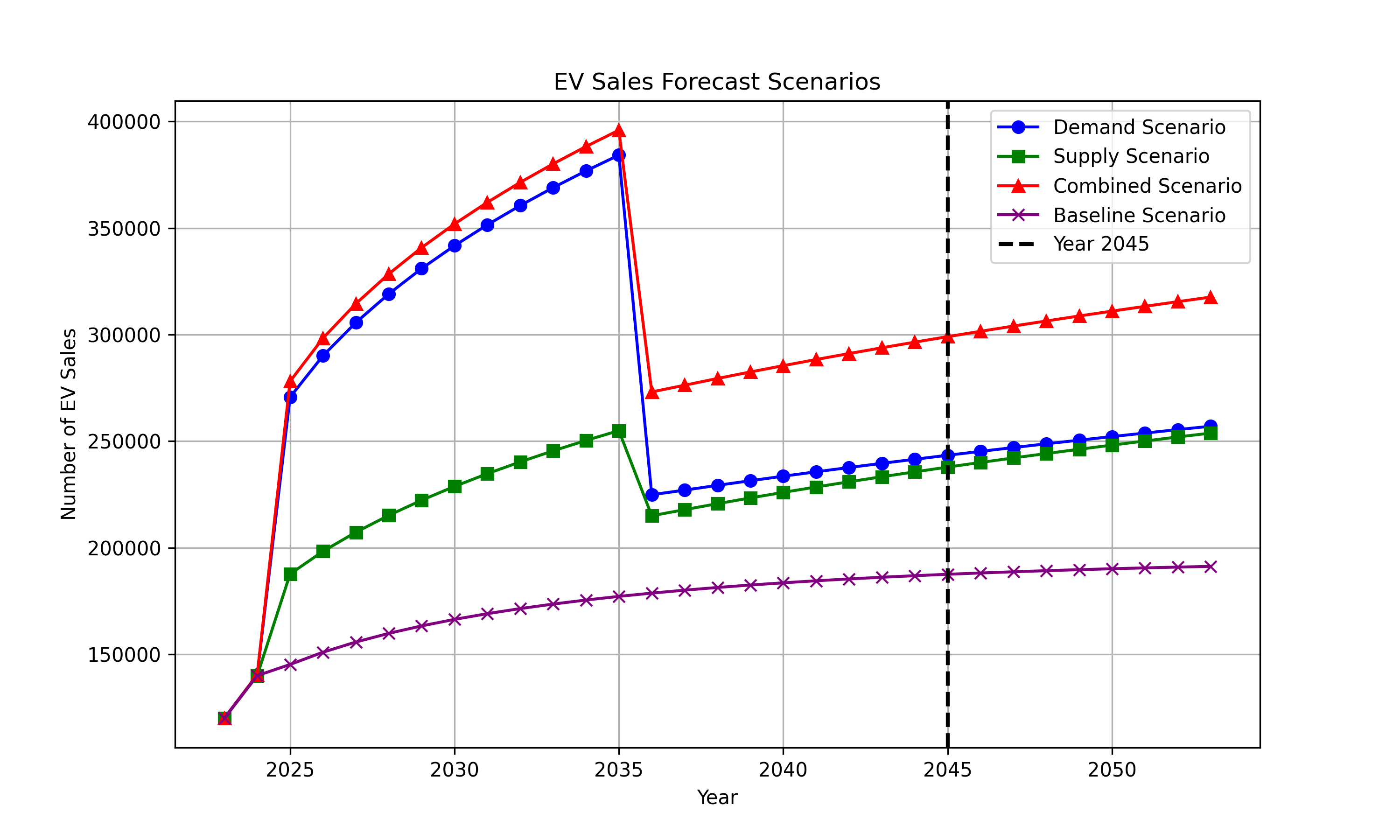}
  \caption{The policy simulation forecasting EV sales based on different magnitudes of policy incentives}
  \label{fig:linechar}
\end{figure}

According to Figure~\ref{fig:linechar}, in our baseline scenario, which reflects the current policies, the forecasted sales of EVs fall below 200,000, resulting in 2.9 million EVs in 2045. In other words, maintaining the current level of incentives will result in less than 50\% of private vehicles being EVs by 2045 (437,000 EVs). This suggests that the current policies are insufficient to stimulate market response. 

To fill this significant gap, we tested several scenarios with stronger incentives to achieve the target. We assume these enhanced incentives would only apply to the next decade, from 2024 to 2035, after which they would return to the baseline level. Empirical studies indicate that policy incentives' effectiveness, especially in economic and social arenas, often has time limitations due to changing economic conditions, the need for policy adaptation, and the dynamic nature of societal responses \cite{c49}. 

We start by isolating the two-sided market, providing incentives for one side to test the effectiveness of direct and indirect network effects. The demand side, which directly influences annual EV sales, is targeted with tax rebates for EV purchases. The current rebate under the Clean Vehicle Rebate Program is \$7,500 for purchasing or leasing a new EV. To reach the 2045 goal, this rebate policy needs to increase by 91\%, leading to EV sales of 375,000 by 2035. By 2045, EVs will make up 73\% of all vehicles.

On the EVCS side, A 100\% EVCS rebate increase yields a less significant but longer-lasting EV sales increase. Per Equations~\eqref{eq:(5)} and~\eqref{eq:(6)}, indirect network effects on EV adoption have a coefficient of $\beta_1\alpha_1$, reflecting both market sides' influence. Figure~\ref{fig:linechar} shows this investment's long-term benefits, with a less significant drop compared to other scenarios. The indirect network effect forms a persistent feedback loop with the demand market. Thus, EVCS investment strengthens this effect and reduces demand drops post-policy. In isolation, the public responds strongly to the direct network effect, while the indirect effect offers sustained benefits through feedback loops.

We then evaluate a scenario that combines the incentives of both sides of the market, effectively leading to 80\% of private vehicles being EVs by 2045. In this scenario, the magnitude of incentives for both market sides is reduced: the EV purchase rebate increases by 60\%, and the public EVCS rebate increases by 66\%. With these amounts for the next decade, EV sales in 2035 can reach around 400,000, and the percentage of EVs on the road can reach 81\% in 2045. The combined scenario leverages both market sides, absorbing high elasticity from the demand side while ensuring acceptable EV adoption levels after policy rescission through the indirect network effect. Since combining the incentives of both sides can effectively reach the goal, policymakers should understand the importance of leveraging both market sides' incentives for a better policy impact.

\section{Discussion} 

Our two-sided model examines EV market dynamics, focusing on EVCS indirect network effects on EV adoption. Using zip code-level data and instrumental variables, we capture nuanced impacts of socio-demographic attributes and market dynamics. Sensitivity analyses and LA County experiments reveal that a 1\% increase in public EVCS leads to a 0.35\% increase in EV sales, while a 1\% increase in total EVs leads to a 0.5\% increase in public EVCS, forming a positive feedback loop. Our forecast suggests that increasing EV purchase rebates by 60\% and EVCS incentives by 66\% over the next decade could achieve 80\% EV adoption for private vehicles by 2045.

Incorporating regional-specific components like EV burden revealed varying responses to policy incentives across zip codes, with higher-burden areas showing lower EV sales compared to higher-income regions. Our policy implications, however, didn't address urban equity issues. Future research should focus on developing more nuanced incentives to promote equitable ZEV transition in disadvantaged communities, expanding our methodology to provide better policy recommendations.

\section{Acknowledgement}

The Center of Excellence on New Mobility and Automated Vehicles (Mobility COE) and the CACLEAN Project support this research.






\bibliographystyle{IEEEtran}
\bibliography{reference}

\end{document}